\documentclass[aps,prd,12pt,notitlepage,showpacs,nofootinbib,tightenlines]{revtex4-1}
\usepackage{amsmath}
\usepackage{amssymb}
\usepackage{epsfig}
\usepackage{graphicx}
\usepackage{bm}
\usepackage{times}
\usepackage{braket}
\usepackage{color}
\usepackage{slashed}
\usepackage{hyperref}
\definecolor{Red}{rgb}{1.,0.,0.}

\definecolor{Blue}{rgb}{0.,0.,1.}

\definecolor{nicered}{rgb}{0.7,0.1,0.1}
\definecolor{nicegreen}{rgb}{0.1,0.5,0.1}
\hypersetup{colorlinks,citecolor=nicegreen,linkcolor=nicered}

\begin{document}
%%%%%%%%%%%%%%%%%%%%%%%%%%%%%%%%%%%%%%%%%%%%%

\newcommand{\beq}{\begin{eqnarray}}
\newcommand{\eeq}{\end{eqnarray}}
\newcommand{\non}{\nonumber\\ }

\newcommand{\jpsi}{J/\Psi}

\newcommand{\ppa}{\phi_\pi^{\rm A}}
\newcommand{\ppp}{\phi_\pi^{\rm P}}
\newcommand{\ppt}{\phi_\pi^{\rm T}}
\newcommand{\ov}{ \overline }

\newcommand{\zerot}{ {\textbf 0_{\rm T}} }
\newcommand{\kt}{k_{\rm T} }
\newcommand{\kta}{{\textbf k_{\rm 1T}} }
\newcommand{\ktb}{{\textbf k_{\rm 2T}} }
\newcommand{\ktc}{{\textbf k_{\rm 3T}} }
\newcommand{\fb}{f_{\rm B} }
\newcommand{\fk}{f_{\rm K} }
\newcommand{\rk}{r_{\rm K} }
\newcommand{\mb}{m_{\rm B} }
\newcommand{\mw}{m_{\rm W} }
\newcommand{\im}{{\rm Im} }

\newcommand{\kks}{K^{(*)}}
\newcommand{\acp}{{\cal A}_{\rm CP}}
\newcommand{\pb}{\phi_{\rm B}}

\newcommand{\xeba}{\bar{x}_2}
\newcommand{\xsba}{\bar{x}_3}
\newcommand{\peas}{\phi^A}

\newcommand{\pvsl}{ p \hspace{-2.0truemm}/_{K^*} }
\newcommand{\esl}{ \epsilon \hspace{-2.1truemm}/ }
\newcommand{\psl}{ p \hspace{-2truemm}/ }
\newcommand{\ksl}{ k \hspace{-2.2truemm}/ }
\newcommand{\lsl}{ l \hspace{-2.2truemm}/ }
\newcommand{\nsl}{ n \hspace{-2.2truemm}/ }
\newcommand{\vsl}{ v \hspace{-2.2truemm}/ }
\newcommand{\epsl}{\epsilon \hspace{-1.8truemm}/\,  }
\newcommand{\bfkk}{{\bf k} }
\newcommand{\calm}{ {\cal M} }
\newcommand{\calh}{ {\cal H} }

%%---------------------------------------------------------
%%%%%%%%%%%%%%%%%%%
\def \appb{{\bf Acta. Phys. Polon. B }  }
\def \cpc{ {\bf Chin. Phys. C } }
\def \ctp{ {\bf Commun. Theor. Phys. } }
\def \epjc{{\bf Eur. Phys. J. C} }
\def \jhep{{\bf J. High Energy Phys. } }
\def \jpg{ {\bf J. Phys. G} }
\def \mpla{{\bf Mod. Phys. Lett. A } }
\def \npb{ {\bf Nucl. Phys. B} }
\def \plb{ {\bf Phys. Lett. B} }
\def \pr{  {\bf Phys. Rep.} }
\def \prc{ {\bf Phys. Rev. C }}
\def \prd{ {\bf Phys. Rev. D} }
\def \prl{ {\bf Phys. Rev. Lett.}  }
\def \ptp{ {\bf Prog. Theor. Phys. }}
\def \zpc{ {\bf Z. Phys. C}  }

%%%%%%%%%%%%%%%%%%%%%%%%%%%%%%%%%%%%%%%%%%%%%%%%%%%%
%%
\title{Time-like pion electromagnetic form factors in $k_{T}$ factorization with the
Next-to-leading-order twist-3 contribution}
\author{Shan Cheng$^{1}$}\email{chengshan-anhui@163.com}
\author{Zhen-Jun Xiao$^{1,2}$}\email{xiaozhenjun@njnu.edu.cn}
\affiliation{1.  Department of Physics and Institute of Theoretical Physics,
Nanjing Normal University, Nanjing, Jiangsu 210023, People's Republic of China,}
\affiliation{2. Jiangsu Key Laboratory for Numerical Simulation of Large Scale Complex Systems,
Nanjing Normal University, Nanjing 210023, People's Republic of China}
\date{\today}
\vspace{1cm}
\begin{abstract}
We calculate the time-like pion electromagnetic form factor in the $k_T$ factorization formalism
with the inclusion of the next-to-leading-order(NLO) corrections to the leading-twist and sub-leading-twist
contributions.
It's found that the total NLO correction can enhance (reduce) the magnitude (strong phase)
of the leading order form factor by $20\% - 30\%$ ( $< 15^o$) in the considered invariant
mass squared $q^2 > 5$ GeV$^2$, and the NLO twist-3 correction play the key role to narrow the gap
between the pQCD predictions and the measured values for the time-like pion electromagnetic form factor.
\end{abstract}

%\vspace{1cm}

%%\keywords{$\kt$ factorization, next-to-leading-order correction, scalar $\pi$ form factors, time-like, B decays.}

\maketitle

\section{Introduction}

As a very important physical observable which may help us to understand the hadrons' structure
and the transition from the perturbative to the non-perturbative region,
the electromagnetic (EM) form factor of the pion meson has been the hot subject of numerous experimental
and theoretical investigations.

During past four decades, the pion EM form factors\cite{prl43-545} have been measured frequently
by many groups \cite{PRD9-1229,npb137-294,prl97-192001,prd8-92,plb073-226,plb220-321,prl95-261803}.
The space-like pion EM form factor was firstly measured by the Harvard $\&$ Cornell collaboration
in the range $0.15 \leq Q^2 \leq 10$ GeV$^2$ of the electro-production processes in 1970s
\cite{PRD9-1229}, and then measured by the DESY collaboration at the fixed point
($Q^2=0.35,~0.70$ GeV$^2$) in the similar processes almost at the same time \cite{npb137-294}.
In the new century, this space-like form factor was determined separately
by the Jefferson Lab $F_{\pi}$ Collaboration
in the region $0.60 \leq Q^2 \leq 1.60$ and at the fixed point $Q^2=2.45$ GeV$^2$\cite{prl97-192001}.
For the time-like pion EM form factor, Cyclotron Laboratory reported their result at the point
$q^2=0.176$ GeV$^2$ in the electro-production process\cite{prd8-92},
then NOVOSIBIRSK collaborations and ORSAY collaboration measured this form factor
independently in the region $0.64 \leq q^2 \leq 1.40$ GeV$^2$\cite{plb073-226}
and $1.35 \leq q^2 \leq 2.38$ GeV$^2$ \cite{plb220-321} through the $e^+ e^-$ annihilation
process respectively.
Recently, CLEO Collaboration also reported their precision measurements of this form factor at the
relatively large $q^2$ ($q^2 = 9.6,~13.48$ GeV$^2$) \cite{prl95-261803}.
A comprehensive summary of experimental measurements for the time-like pion EM form factor
can be found in Ref.~\cite{jpg29-A1}.

On the theory side, pion EM form factors also attracted much attentions.
The space-like one was studied at different energy regions in QCD by using the different approaches.
For example, it was investigated in high and intermediate energy region in
Ref.~\cite{prl65-1717} and Refs.~\cite{prd61-073004,jhep07-008} respectively,
while it's asymptotic behavior at the extremely large $q^2$ was studied in Ref.~\cite{plb094-245}.
In Refs.~\cite{prd60-074004,prd70-033014,prd79-034015}, the space-like pion form factor was
studied carefully in the perturbative QCD theory, and it was also studied in the sum rules
formalism\cite{plb115-410}.
For the time-like pion EM form factor\cite{plb316-546},
it's high $q^2$ behavior was determined at $q^2 = M^2_{J/\Psi}$ and it was found that
it is larger than the space-like one by a factor of 2 \cite{prd47-R3690}.
The Sudakov effect for the time-like form factor was discussed in
Refs.~\cite{prd51-15,prd62-113001}
and it's found that the asymptotic behavior of the integrable singularity for the time-like form factor
is the same as that for the space-like one.
The conformal symmetry was also used to analyze the time-like form factor\cite{prd77-113004}
and it is shown explicitly that the time-like form factor, which was obtained by the
analytic continuation of the space-like one, satisfied correctly the dispersion relation.
The light-cone QCD investigation recently\cite{prd82-114012,plb693-102} showed that the effects
of the power suppressed sub-leading twist' and the genuine soft QCD correction' contributions
turn out to be dominant at low- and moderate-energies.

With removing the end-point singularities by the Sudakov factors \cite{prd47-3875,prd66-094010},
the $k_T$ factorization theorem\cite{kt theorem} is successful to deal with the exclusive processes
with a large momentum transfer\cite{prd67-034001}.
In the $k_T$ factorization theorem, the space-like pion EM form factor was re-examined
with the inclusion of the Sudakov suppression \cite{prd52-5358}.
Three-parton contribution to pion EM form factor in $k_T$ factorization was also
investigated in Refs.~\cite{prd84-034018} and it's found that such contribution is rather small
in size and therefore can be dropped safely.

After completing the NLO calculations for the space-like pion EM form factor at leading twist (twist-2)
\cite{prd83-054029}, the authors also studied the NLO twist-2 time-like pion EM form factor
~\cite{plb718-1351} and found that the NLO correction to the  LO magnitude (strong phase) is lower than
$25\%$ ( $10^o$) at the large invariant mass squared $q^2 > 30$ GeV$^2$ at leading twist.
In Refs.~\cite{prd63-074009,prd89-054015}, the authors calculated the sub-leading twist's (twist-3)
contribution from pion meson distribution amplitudes(DAs) to the exclusive $B \to \pi$ transition
form factors and the space-like pion EM form factor, and they found that
this power-suppressed contribution is large in the low and intermediate $q^2$ regions.
In this paper, therefore, we will evaluate the NLO twist-3 contribution to the time-like pion
EM form factor after the calculation for the NLO correction to the space-like pion EM form factor
at twist-3 level \cite{prd89-054015}.

This paper is organized as follows.
In Sec.~II, we give the Leading order(LO) analysis for the time-like pion EM form factor.
In Sec.~III, the NLO twist-3 corrections to the time-like form factor will be calculated
from the space-like one by analytical continuation.
Sec.~IV contains the numerical analysis of the NLO effects, and the conclusion
will also be given in this section.

%%%%%%%%%%%%%%%%%%%%%%%%%%%%%%%%%%%%%%%%%%%%%%%
\section{LO analysis}
%%-----------------------------------------------------------------------
\begin{figure}[tb]
\vspace{-2cm}
\begin{center}
\leftline{\epsfxsize=16cm\epsffile{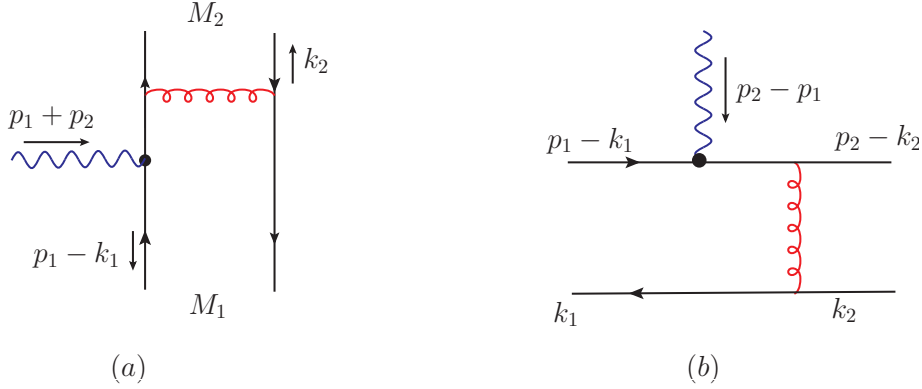}}
\end{center}
\vspace{-16.5cm}
\caption{The LO diagrams for the time-like (a) and space-like (b) pion electromagnetic form factor,
with $\bullet$ here representing the electromagnetic vertex.}
\label{fig:fig1}
\end{figure}
%%-----------------------------------------------------------------------

In this section we present the LO factorization formula for the time-like pion EM form factor
and evaluate  the contributions from the two-parton twist-2 and twist-3 pion meson DAs.
The LO quark diagram for the relative time-like and space-like pion EM form factor corresponding
to the process $\gamma^{\star} \to \pi \pi$($\pi \gamma^{\star}\to \pi$ ) are illustrated
in Fig.~\ref{fig:fig1}(a) and \ref{fig:fig1}(b), respectively.

One should note that the kinetics for the time-like pion EM form factor are different from that
for the space-like form factor, because the two mesons are both outgoing in Fig.~\ref{fig:fig1}(a),
but one is incoming and another is outgoing in Fig.~\ref{fig:fig1}(b).
In the light-cone coordinates, the momenta $p_1$ and $p_2$ in Fig.~\ref{fig:fig1}(a) are
parameterized as
\beq
p_1&=&(p^+_1,0,\mathbf{0_T}),~~~~~p_2=(0,p^-_2\mathbf{,0_T});~~~~~p^+_1=p^-_2=\frac{Q}{\sqrt{2}},
\label{eq:kinetic1} \\
k_1&=&(x_1p^+_1,0,\mathbf{k}_{1T}),~~~~~k_2=(0,x_2p^-_2,\mathbf{k}_{2T}),~~~~~q^2=Q^2=(p_1+p_2)^2,
\label{eq:kinetic2}
\eeq
with $q^2$ being the invariant mass squared of the intermediate virtual photon,
$k_1$ ($k_2$) is the momentum carried by the valence quark (anti-quark) of meson $M_1$ ($M_2$)
with the momentum fraction $x_1$ ($x_2$) denoting the strength of the quark (anti-quark)
to form the corresponding meson.
Then the time-like (space-like) pion EM form factor $G_{\pi}$ ($F_{\pi}$) can be specified
through the following matrix elements\cite{plb693-102}:
\beq
e(p_1-p_2)_{\mu}G_{\pi}(q^2)&=&<\pi^{\pm}(p_2) \pi^{\mp}(p_1) \mid J^{EM}_{\mu}(p_1+p_2) \mid 0>,
\label{eq:define-tff} \\
e(p_1+p_2)_{\mu}F_{\pi}(Q^2)&=&<\pi^{\pm}(p_2) \mid J^{EM}_{\mu}(p_1-p_2) \mid \pi^{\pm}(p_1)>,
\label{eq:define-sff}
\eeq
where $J^{EM}_{\mu}$ is the EM current.
The space-like momentum transfers in Eq.~(\ref{eq:define-sff}) is $Q^2=-q^2=-(p_1-p_2)^2$,
which is different from  the time-like one as described in Eq.~(\ref{eq:kinetic2}).

From Fig.~\ref{fig:fig1}(a) and Fig.~\ref{fig:fig1}(b), one can write down the LO time-like and
space-like hard kernels
\beq
H^{(0)}_{a}(x_i, \mathbf{k_{iT}}, Q^2) &=& \frac{- i e_q 32 \pi \alpha_s C_F N_C Q^2}{(p_2+k_1)^2 (k_2+k_1)^2} \cdot
           \Big \{ x_1p_{1\mu} \phi^A(x_1) \phi^A(x_2) \non
           && - 2 r^2_0 \left [(p_{2\mu} + x_1p_{1\mu}) \phi^P(x_1) \phi^P(x_2)
                        -(p_{2\mu} - x_1p_{1\mu}) \phi^T(x_1) \phi^P(x_2)\right ] \Big \}, \label{eq:lohka}\\
H^{(0)}_{b}(x_i, \mathbf{k_{iT}}, Q^2) &=& \frac{i e_q 32 \pi \alpha_s C_F N_C Q^2}{(p_2-k_1)^2 (k_2-k_1)^2} \cdot
           \Big \{ x_1p_{1\mu} \phi^A(x_1) \phi^A(x_2) \non
           &&+ 2 r^2_0 \left [(p_{2\mu} - x_1p_{1\mu}) \phi^P(x_1) \phi^P(x_2)
                       -(p_{2\mu} + x_1p_{1\mu}) \phi^T(x_1) \phi^P(x_2)\right ] \Big \},
\label{eq:lohkb}
\eeq
where $\phi^{A}(x_1)$ and $\phi^{P,T}(x_1)$ ($\phi^{A}(x_2)$ and $\phi^{P,T}(x_2)$) represent
the twist-2 and twist-3 pion meson DAs for the corresponding meson with the momentum $p_1$ ( $p_2$ ),
the chiral parameter is defined as $r^2_0=m^2_0/Q^2$ with the chiral mass $m_0=1.74$ GeV.
By comparing Eq.~(\ref{eq:lohka}) with Eq.(\ref{eq:lohkb}), we can find that the LO
time-like hard kernel has the similar structure with that for the space-like one,
the only difference is the direction of the valence quark momentum $k_1$, which will flow into the
internal propagators.
Then we can obtain the LO time-like hard kernel from the space-like one by direct replacement
$-\ksl_1 \to \ksl_1$ for the internal propagators, which implied that the time-like form factor
can also be obtained from the space-like one by analytical continuation from $-Q^2$ to $Q^2$
in the invariant mass squared $q^2$ space.
This is the basic idea being used to calculate the NLO time-like pion EM form factor in this paper.

The LO time-like and space-like pion EM form factor can be obtained by combining
Eqs.~(\ref{eq:define-tff},\ref{eq:lohka})
and Eqs.~(\ref{eq:define-sff},\ref{eq:lohkb}) respectively, and can then be written in the
following forms:
\beq
Q^2G^{(0)}(x_i,Q^2,\textbf{k}_{iT}) &= & \frac{128 \pi Q^4 \cdot \alpha_s(\mu)}{(p_2+k_1)^2(k_1+k_2)^2}
\int^1_0 dx_1 dx_2 \int^{\infty}_0 \frac{d^2\textbf{k}_{1T}}{2\pi} \frac{d^2\textbf{k}_{2T}}{2\pi}\non
&&\hspace{-2.5cm} \cdot \Big \{ -x_1 \phi^A(x_1) \phi^A(x_2)
 + 2 r^2_0 \left [(1-x_1) \phi^P(x_1) \phi^P(x_2) + (1+x_1) \phi^T(x_1) \phi^P(x_2) \right ] \Big \},
\label{eq:lotff-kt}\\
Q^2F^{(0)}(x_i,Q^2,\textbf{k}_{iT}) &= & \frac{128 \pi Q^4 \cdot \alpha_s(\mu)}{(p_2-k_1)^2(k_1-k_2)^2}
\int^1_0 dx_1 dx_2 \int^{\infty}_0 \frac{d^2\textbf{k}_{1T}}{2\pi} \frac{d^2\textbf{k}_{2T}}{2\pi}\non
&&\hspace{-2.5cm}\cdot \Big \{ x_1 \phi^A(x_1) \phi^A(x_2)
 + 2 r^2_0 \left [(1-x_1) \phi^P(x_1) \phi^P(x_2) - (1+x_1) \phi^P(x_1) \phi^T(x_2) \right ] \Big \}.
\label{eq:losff-kt}
\eeq
The relation $\phi^T(x)=\partial\phi^A(x)/6\partial_x$ has been considered
in the process to derive out Eq.~(\ref{eq:lotff-kt}).

For the time-like case, the denominator in Eq.~(\ref{eq:lotff-kt}) is expanded as
\beq
(p_2+k_1)^2(k_1+k_2)^2 = (x_1Q^2 - \mathbf{k}^2_{1T} + i \epsilon)
                         (x_1x_2Q^2 - \mid \mathbf{k}_{1T} + \mathbf{k}_{2T} \mid^2 + i \epsilon),
\label{eq:denominator-lohka}
\eeq
and then the internal gluon/quark may go on mass shell, which will generate an image part in the
hard kernel according to the principle-value prescription:
\beq
\frac{1}{k^2_{T} - \beta -i \epsilon} = Pr \frac{1}{k^2_{T} - \beta} + i \pi \cdot \delta(k^2_{T} - \beta).
\label{eq:principle-value}
\eeq
But in the space-like case, no image part would appeared because the internal gluon/quark can't go on
mass shell because the denominator in Eq.~(\ref{eq:lohkb}) should be expanded as
\beq
(p_2-k_1)^2(k_1-k_2)^2 = (x_1Q^2 + \mathbf{k}^2_{1T})(x_1x_2Q^2 + \mid \mathbf{k}_{1T} + \mathbf{k}_{2T} \mid^2).
\label{eq:denominator-lohkb}
\eeq

Now we consider the end-point behaviors of the LO form factors.
For the elaboration, we here show the end-point behaviors in
Eqs.~(\ref{eq:lotff-kt},\ref{eq:losff-kt}) by using the asymptotic pion meson DAs only \cite{prd89-054015}:
\beq
\phi_{\pi}^{A}(x) = 6 f_{\pi} x(1-x),  \quad
\phi_{\pi}^{P}(x) = f_{\pi}, \quad
\phi_{\pi}^{T}(x) = f_{\pi} (1-2x).
\label{eq:DAs-asy}
\eeq
Then the end-point behaviours of the integrands in Eqs.~(\ref{eq:lotff-kt},\ref{eq:losff-kt}) can be expressed roughly as
\beq
Q^2G^{(0)}(x_i,Q^2,\textbf{k}_{iT}) &\varpropto&
\frac{- 9 x_1 x_1 x_2 (1 - x_1) (1 - x_2) + r^2_{\pi} (1 - x_1 - x^2_1)}{(p_2+k_1)^2(k_1+k_2)^2},
\label{eq:end-point-lotff-kt}\\
Q^2F^{(0)}(x_i,Q^2,\textbf{k}_{iT}) &\varpropto& \frac{ 9 x_1 x_1 x_2 (1 - x_1) (1 - x_2) + r^2_{\pi} x^2_1}{x_1 x_1 x_2 Q^4},
\label{eq:end-point-losff-kt}
\eeq
where the first(second) terms in Eqs.~(\ref{eq:end-point-lotff-kt},\ref{eq:end-point-losff-kt})
are the contributions arose from twist-2(twist-3) DAs.
In the expansions of Eq.~(\ref{eq:end-point-losff-kt}), the transverse momentum contributions
in the internal propagators was absorbed into the effective momentum fraction $x_i$.
From the expressions in Eqs.~(\ref{eq:lotff-kt}-\ref{eq:end-point-losff-kt}), one can see
the following points:
\begin{enumerate}
\item [(i)]
The contribution to the LO pion EM form factor from the twist-2 DAs,
no matter for the time-like case or the space-like one,
has no end-point singularity because of the cancelation of them
between the denominator and numerator.
The contribution arose from the twist-3 DAs, however, will generate the end-point
singularities, although they are power-suppressed by $r^2_{\pi}$ in the large momentum
transfers region.
The LO space-like form factor from the twist-3 DAs is behaved as $1/x_1$, for example,
the twist-3 DAs would give the dominate contribution in the small and intermediate
momentum transfers region.

\item [(ii)]
Since the Sudakov factor from threshold resummation\cite{prd66-094010}
can suppress effectively the end-point singularity from the twist-3 contribution,
a rough estimate shows that the major contribution to the LO space-like form factor in
Eq.~(\ref{eq:end-point-losff-kt}) comes from the region of $x_1 \sim 0.1$ and $x_2 \sim 0.5$.
Then the contribution to the LO space-like form factor from the twits-2 DAs will become
as large as that from the twist-3 DAs at the point $Q^2 \sim 7.4$ GeV$^2$, which has been
confirmed by the numerical result in  Ref.~\cite{prd89-054015}.

\item [(iii)]
The second terms in Eq.~(\ref{eq:end-point-lotff-kt}) is proportional to $1-x_1-x^2_1$,
which is much larger than the second term  in Eq.~(\ref{eq:end-point-losff-kt}) since this second
term is proportioned to $x^2_1\sim 10^{-2}$.
The end-point singularity for the time-like form factor induced by the twist-3 DAs,
consequently, is much higher than that for the space-like one.
The twist-3 contribution to the time-like form factor is then much larger than
the twist-2 contribution in the low and intermediate $q^2$ region.
Simple estimation shows that these two kinds of contributions may become similar
in size in the high $Q^2 \sim 300$ GeV$^2$ region.

\end{enumerate}

By making the Fourier transformation for function $Q^2G^{(0)}(x_i,Q^2,\textbf{k}_{iT})$ in
Eq.~(\ref{eq:lotff-kt}) from the transversal momentum space ($\mathbf{k}_{iT})$
to the conjugate-parameter space ($\mathbf{b}_{i}$), we obtain the standard double-b convolution
LO time-like pion EM form factor\cite{plb693-102,plb718-1351,npb2015}:
\beq
Q^2G^{(0)}_{II} &&= \int^1_0 dx_1 dx_2 \int^{\infty}_0 b_1 db_1 b_2 db_2 ~ 128 \pi Q^4
\cdot \alpha_s(\mu) \cdot \textmd{exp}[-S(x_i;b_i;Q;\mu)] \non
&& \cdot \{ - x_1 \phi^A(x_1) \phi^A(x_2) + 2 r^2_0 \left[(1-x_1)\phi^P(x_1)\phi^P(x_2)+(1+x_2)\phi^T(x_1)\phi^P(x_2) \right] \cdot S_t(x_i) \} \non
&& \cdot K_0(i \sqrt{x_1 x_2} Q b_2)\cdot \left[ K_0(\sqrt{x_1} Q b_1) I_0(\sqrt{x_1} Q b_2) \theta(b_1-b_2) + (b_1 \leftrightarrow b_2) \right],
\label{eq:lotff-double}
\eeq
where the Sudakov exponent $S=S(x_1,b_2;M_B;\mu)+S(x_2,b_2;M_B;\mu)$
is the $k_T$ resummation factor,
the Sudakov factor $S_t(x_i)=S_t(x_1) \cdot S_t(x_2)$ refers to the threshold resummation factor,
$K_0$ and $I_0$ are the Bessel functions:
\beq
K_0(iz)=\frac{i\pi}{2} H^{(1)}_0(iz); ~~H^{(1)}_0(iz)=H^{(1)}_0(z)=J_0(z)+iN_0(Z);  ~~I_0(z)=J_0(z).
\label{eq:Besell}
\eeq

Since the $k_T$ factorization theorem applies to processes dominated by small $x$ contribution,
so the NLO correction to the space-like pion EM form factor
\cite{prd83-054029,prd89-054015} has been calculated
with the hierarchy $x_1Q^2,x_2Q^2 \gg x_1x_2Q^2, \mathbf{k}^2_T$ for convenience.
Since there is no end-point singularity for the LO pion form factor from the twist-2 DAs,
we can ignore the transverse momenta for the internal quark propagator safely
for the twist-2 contribution as elaborated in Ref.~\cite{plb718-1351},
then the denominator for the first term in Eq.~(\ref{eq:lotff-kt}) is reduced to
\beq
(p_2+k_1)^2(k_1+k_2)^2 = x_1Q^2 (x_1x_2Q^2 - \mid \mathbf{k}_{1T} + \mathbf{k}_{2T} \mid^2 + i \epsilon).
\label{eq:denominator-lohka-I}
\eeq
The LO time-like pion EM form factor from the twist-2 DAs can be written in a
single-b convolution formula as follows:
\beq
Q^2G^{(0)}_{T2,I} &&= \int^1_0 dx_1 dx_2 \int^{\infty}_0 b_1 db_1 b_2 db_2 ~ 128 \pi Q^4
\cdot \alpha_s(\mu) \cdot \textmd{exp}[-S(x_i;b_i;Q;\mu)] \non
&& \cdot \Big \{ - x_1 \phi^A(x_1) \phi^A(x_2) \Big \} \cdot K_0(i \sqrt{x_1 x_2} Q b_2).
\label{eq:lotff-single}
\eeq
In Ref.~\cite{plb718-1351}, the authors confirmed that the numerical results of form factor in
the standard double-b convolution of  Eq.~(\ref{eq:lotff-double}) is approximately equal to the
value of single-b convolution of Eq.~(\ref{eq:lotff-single}),
which furthermore showed that the major source of the strong phase is produced by the
internal gluon propagator for the twist-2 contribution.

The LO time-like form factor from the twist-3 DAs, however, has a high power end-point singularity,
the single-b approximation is therefore not valid for the twist-3 DAs's contribution.
So in the next section we have to calculate the NLO twist-3 hard kernel in time-like form factor
by using the double-b convolution method.

\section{NLO correction to the twist-3 time-like pion EM form factors}

The LO analysis in the last section show that the time-like hard kernel can be obtained from
the space-like one by the simple space transfer: $-Q^2 \rightarrow Q^2$.
Because of the Lorentz invariant QCD theory, it's believed that this analytical
continuation should be hold at NLO.

In $k_T$ factorization theorem, the NLO hard kernel for pion EM form factor is derived by taking
the difference of the NLO (${\cal O}(\alpha^2_s)$) quark diagrams and the convolutions of the
LO (${\cal O}(\alpha_s)$) hard kernel with the NLO (${\cal O}(\alpha_s)$) effective diagrams
for meson wave functions.
For the space-like pion EM form factors as described explicitly in Refs.~\cite{prd83-054029,prd89-054015},
the ultraviolet divergences are just absorbed into the renormalized coupling constant $\alpha_s(\mu)$
with the massless pion meson, the infrared divergences arose from the soft region are canceled
by themselves in the quark diagrams, and the infrared divergences in the collinear region
for the quark diagrams can be absorbed into the high order non-perturbative meson wave functions.

For the time-like form factor, the NLO twist-2 hard kernel has been calculated in Ref.~(\cite{plb718-1351})
and then the only unknown NLO correction at present is the one from the twist-3 DAs.
With the NLO twist-3 space-like hard kernels calculated in Ref.~(\cite{prd89-054015}),
we can obtain the NLO twist-3 time-like hard kernel by the analytical continuation $-Q^2 \rightarrow Q^2$.
For this purpose, we firstly define two types of LO twist-3 time-like hard kernels $H^{(0)}_{T3,1}$ ($H^{(0)}_{T3,2}$)
proportioned to the lorentz structure $p_{1 \mu}$ ($p_{2 \mu}$) from Eq.~(\ref{eq:lohka}):
\beq
H^{(0)}_{T3,1}(x_i, \mathbf{k_{iT}}, Q^2) &&= \frac{ i e_q 32 \pi \alpha_s C_F N_C Q^2}{(p_2+k_1)^2 (k_2+k_1)^2} \cdot
            2 r^2_0 x_1p_{1\mu} \left [\phi^P(x_1) + \phi^T(x_1) \right ] \phi^P(x_2),
\label{eq:lohkt3-1} \\
H^{(0)}_{T3,2}(x_i, \mathbf{k_{iT}}, Q^2) &&= \frac{ i e_q 32 \pi \alpha_s C_F N_C Q^2}{(p_2+k_1)^2 (k_2+k_1)^2} \cdot
            2 r^2_0 p_{2\mu} \left [\phi^P(x_1) - \phi^T(x_1) \right ] \phi^P(x_2).
\label{eq:lohkt3-2}
\eeq

By substituting $Q^2 +i \epsilon$ for the momentum transfers of the virtual photon, and
$x_1x_2Q^2 -( \mathbf{k}_{1T}+\mathbf{k}_{1T})^2 + i \epsilon$($x_1Q^2 - \mathbf{k}^2_{1T} + i \epsilon$)
for the internal gluon(quark), we can obtain the NLO twist-3 hard kernels for the time-like $\pi^+\pi^-$
production process from the NLO twist-3 space-like one\cite{prd89-054015}.
The NLO twist-3 time-like hard kernels can then be written as the form of
\beq
H^{(1)}_{T3,1}(x_i, \mathbf{k_{iT}}, Q^2, \mu, \mu_f) &=& h_{T3,1}(x_i,\mathbf{k}_{iT},Q,\mu,\mu_f) \cdot H^{(0)}_{T3,1}(x_i, \mathbf{k_{iT}}, Q^2)
\label{eq:nlohk1} \\
H^{(1)}_{T3,2}(x_i, \mathbf{k_{iT}}, Q^2, \mu, \mu_f) &=& h_{T3,2}(x_i,\mathbf{k}_{iT},Q,\mu,\mu_f) \cdot H^{(0)}_{T3,2}(x_i, \mathbf{k_{iT}}, Q^2).
\label{eq:nlohk2}
\eeq
By setting the renormalized and factorized scales both at the internal hard scale $\mu = \mu_f = t$,
and using the follow relations,
\beq
&&\ln(-Q^2-i\epsilon) = \ln(Q^2) -i\pi, \non
&&\ln(\mathbf{k}^2_{1T} - x_1Q^2 + i \epsilon) = \ln (\mathbf{k}^2_{1T} - x_1Q^2) + i\pi \Theta(\mathbf{k}^2_{1T} - x_1Q^2) \non
&&\ln(\mathbf{k}^2_{T} -x_1x_2Q^2 + i \epsilon) =  \ln (\mathbf{k}^2_{T} -x_1x_2Q^2) + i\pi \Theta(\mathbf{k}^2_{T} -x_1x_2Q^2).
\label{eq:relations-log}
\eeq
the relevant correction functions $h_{T3,1}, h_{T3,2}$ in Eqs.~(\ref{eq:nlohk1},\ref{eq:nlohk2}) can be written as,
\beq
h_{T3,1}(x_i,\mathbf{k}_{iT},Q,t) = \frac{\alpha_s C_F}{4\pi} && \left[ \frac{9}{4}\ln\left (\frac{t^2}{Q^2}\right )
                 - \frac{53}{16}\ln\delta'_{12} - \frac{23}{16}\ln x'_1 - \frac{1}{8}\ln^2x_2 \right. \non
        &&\left.  - \frac{9}{8}\ln x_2 - \frac{137\pi^2}{96} + \frac{337}{64} + i\pi\frac{5}{2} \right],
\label{eq:nlocf1}
\eeq
\beq
h_{T3,2}(x_i,\mathbf{k}_{iT},Q,t) = \frac{\alpha_s C_F}{4\pi} && \left[ \frac{9}{4}\ln\left (\frac{t^2}{Q^2}\right )
                 - 4\ln\delta'_{12} - \frac{1}{2} \ln^2x'_1 + 2\ln x_2 \right. \non
        &&\left.  - \frac{15\pi^2}{24} + \frac{\ln2}{4} + \frac{11}{2} + i\pi\left (\frac{7}{4} + \ln x'_1\right ) \right],
\label{eq:nlocf2}
\eeq
where $\ln\delta'_{12} \equiv \ln( (\mathbf{k}_{1T} + \mathbf{k}_{2T})^2 -x_1x_2Q^2 + i \epsilon) - \ln Q^2$ and
$\ln x'_1 \equiv \ln(\mathbf{k}^2_{1T} - x_1Q^2 + i \epsilon)$.

We can then obtain the NLO twist-3 time-like correction functions $h_{T3,1}, h_{T3,2}$ in the parameter space $\mathbf{b}_i$
by the Fourier transformation from the transverse momentum space $\mathbf{k}_{iT}$ to $\mathbf{b}_i$ space.
The correction functions in $\mathbf{b}$ space takes the form of
\beq
h_{T3,1}(x_i,b_i,Q,t) = \frac{\alpha_s C_F}{4\pi} && \left[ \frac{9}{4}\ln\left (\frac{t^2}{Q^2}\right )
                 - \frac{53}{32}\ln\left (\frac{4x_1x_2}{Q^2b_2^2}\right )
                 - \frac{23}{32}\ln\left (\frac{4x_1}{Q^2b^2_1}\right )
                 - \frac{1}{8}\ln^2x_2 \right. \non
        &&\left.  - \frac{9}{8}\ln x_2 - \frac{137\pi^2}{96} + \frac{19}{4}\gamma_E + \frac{337}{64} + i\pi\frac{39}{8} \right],
\label{eq:nlocf1-b} \\
h_{T3,2}(x_i,b_i,Q,t) = \frac{\alpha_s C_F}{4\pi} && \left[ \frac{9}{4}\ln\left (\frac{t^2}{Q^2}\right )
                 - 2\ln\left (\frac{4x_1x_2}{Q^2b_2^2}\right )
                 - \frac{1}{8} \ln^2\left (\frac{4x_1}{Q^2b^2_1}\right )\right. \non
         &&\left. + \left (\frac{\gamma_E}{2} + \frac{3}{4} i\pi\right )\ln \left (\frac{4x_1}{Q^2b^2_1}\right )
         + 2\ln x_2 - \frac{\pi^2}{4} - \frac{\gamma^2_E}{2} + 4\gamma_E \right.\non
         && \left. + \frac{\ln2}{4}
                 + \frac{11}{2} + i\pi\left (\frac{15}{4} - \frac{3}{2}\gamma_E\right ) \right ],
\label{eq:nlocf2-b}
\eeq
where $\gamma_E$ is the Euler constant.

With the NLO twist-3 correction function in Eqs.~(\ref{eq:nlocf1-b},\ref{eq:nlocf2-b}) and
the NLO twist-2 correction function in Ref.~(\cite{plb718-1351}),
we can derive the NLO time-like pion EM form factor in $k_T$ factorization formula as,
\beq
Q^2G^{(1)}_{\rm II} &&=  128 \pi Q^4 \cdot \alpha_s(\mu) \cdot
\int^1_0 dx_1 dx_2 \int^{\infty}_0 b_1 db_1 b_2 db_2 \cdot \textmd{exp}[-S(x_i;b_i;Q;\mu)] \non
&& \cdot \Big \{ - x_1 \phi^A(x_1) \phi^A(x_2) \cdot h_{T2}
+ 2 r^2_0 \left [ \left (\phi^P(x_1)+\phi^T(x_1) \right )\phi^P(x_2) \cdot h_{T3,2} \right. \non
&&\left. + x_1\left (\phi^T(x_1)-\phi^P(x_1) \right )\phi^P(x_2) \cdot h_{T3,1} \right]
\cdot S_t(x_i) \Big \} \cdot K_0(i \sqrt{x_1 x_2} Q b_2)\non
&&\cdot \left[ K_0(\sqrt{x_1} Q b_1) I_0(\sqrt{x_1} Q b_2) \theta(b_1-b_2) + (b_1 \leftrightarrow b_2) \right],
\label{eq:nlotff-double}
\eeq
where the NLO twist-2 correction function $h_{T2}$ derived from sing-b formula is expressed as
the following form \cite{plb718-1351},
\beq
h_{T2}(x_i, && b_2,Q,t) =\frac{\alpha_s C_F}{4\pi} \Big \{ -\frac{3}{4}\ln\left (\frac{t^2}{Q^2} \right )
                  - \frac{1}{4}\ln^2\left (\frac{4x_1x_2}{Q^2b_2^2}\right )
                  - \frac{17}{4}\ln^2x_1 +\frac{27}{8}\ln x_1\ln x_2 \non
                &&+\left (\frac{17}{8}\ln x_1 + \frac{23}{16} + \gamma_E + i\frac{\pi}{2}\right )
                \ln\left (\frac{4x_1x_2}{Q^2 b_2^2}\right )
                  -\left (\frac{13}{8}+\frac{17\gamma_E}{4}-i\frac{17\pi}{8}\right )\ln x_1 \non
                &&+ \frac{31}{16}\ln x_2 -\frac{\pi^2}{2} + (1-2\gamma_E)\pi + \frac{\ln2}{2}
                  + \frac{53}{4} - \frac{23\gamma_E}{8} - \gamma_E^2 + i\pi \left (\frac{171}{16}+\gamma_E \right ) \Big \}.
\label{eq:nlocf-t2}
\eeq

\section{Numerical results and discussions}

In this section we present the numerical results for the time-like pion EM form factor  induced by
the distribution amplitudes with different twists at LO and NLO level.
Non-asymptotic pion meson DAs as given in Eq.~(\ref{eq:DAs-non-asy}) with the inclusion
of the high order effects are adopted in our numerical calculation.
\beq
\phi_{\pi}^{A}(x) &= &\frac{3 f_{\pi}}{\sqrt{6}} x (1-x)
\left [ 1 +  a_2^{\pi} C_2^{\frac{3}{2}}(u) + a_4^{\pi} C_4^{\frac{3}{2}}(u) \right ], \non
\phi_{\pi}^{P}(x) &= &\frac{f_{\pi}}{2\sqrt{6}} \left[ 1 + \left (30 \eta
_3 - \frac{5}{2} \rho_{\pi}^2 \right ) C_2^{\frac{1}{2}}(u) - 3 \left (\eta_3 \omega_3
+ \frac{9}{20} \rho_{\pi}^2 \left (1 + 6 a_2^{\pi}\right ) \right ) C_4^{\frac{1}{2}}(u) \right ],  \non
\phi_{\pi}^{T}(x) &= & \frac{f_{\pi}}{2\sqrt{6}} (1-2x)
\left [1 + 6 \left (5 \eta_3 - \frac{1}{2} \eta_3 \omega_3 - \frac{7}{20} \rho_{\pi}^2
- \frac{3}{5} \rho_{\pi}^2 a_2^{\pi} \right ) \left (1-10 x + 10 x^2 \right )  \right ],
\label{eq:DAs-non-asy}
\eeq
where the Gegenbauer moments $a_i^\pi$, the parameters $\eta_3, \omega_3$ and $\rho_\pi$ are adopted from
Refs.~\cite{prd49-2525,jhep01-010,jhep05-004,prd71-014015}:
\beq
a_2^{\pi} &=& 0.25, \quad a_4^{\pi} = -0.015, \quad
\rho_{\pi} = m_{\pi}/m_0, \quad \eta_3 = 0.015, \quad \omega_3 = -3.0,
\label{eq:input1}
\eeq
with $f_\pi=0.13$ GeV, $m_\pi=0.13$ GeV,  $m_0=1.74$ GeV.

%-----------------------------------------------------------------------
\begin{figure*}
\centering
\vspace{-0.5cm}
\includegraphics[width=0.48\textwidth]{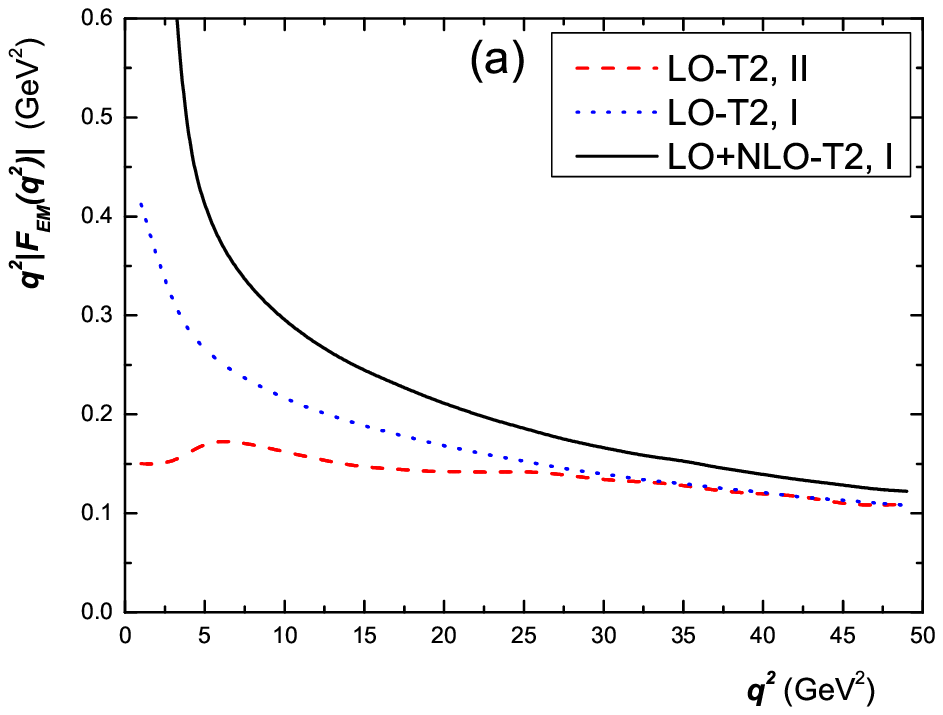}
\includegraphics[width=0.48\textwidth]{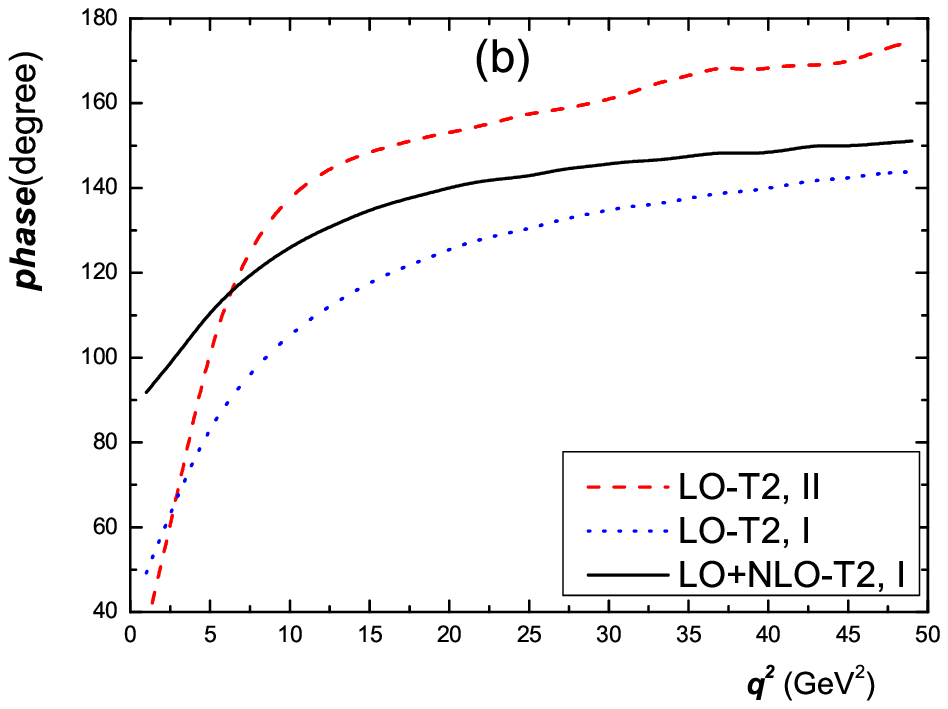}
\vspace{-0.5cm}
\caption{The pQCD predictions for the magnitude and strong phase of time-like pion EM form factors
induced by the twist-2 DAs $\phi^A$.
The Rome symbol ``II"(``I") refers to the form factors calculated in double-b(single-b)
convolution formula as described in Eq.~(\ref{eq:lotff-double}) ( Eq.~(\ref{eq:lotff-single}) ).}
\label{fig:fig2}
\end{figure*}
%%------------------------------------------------------------------------

%-----------------------------------------------------------------------
\begin{figure*}
\centering
\vspace{-0.5cm}
\includegraphics[width=0.48\textwidth]{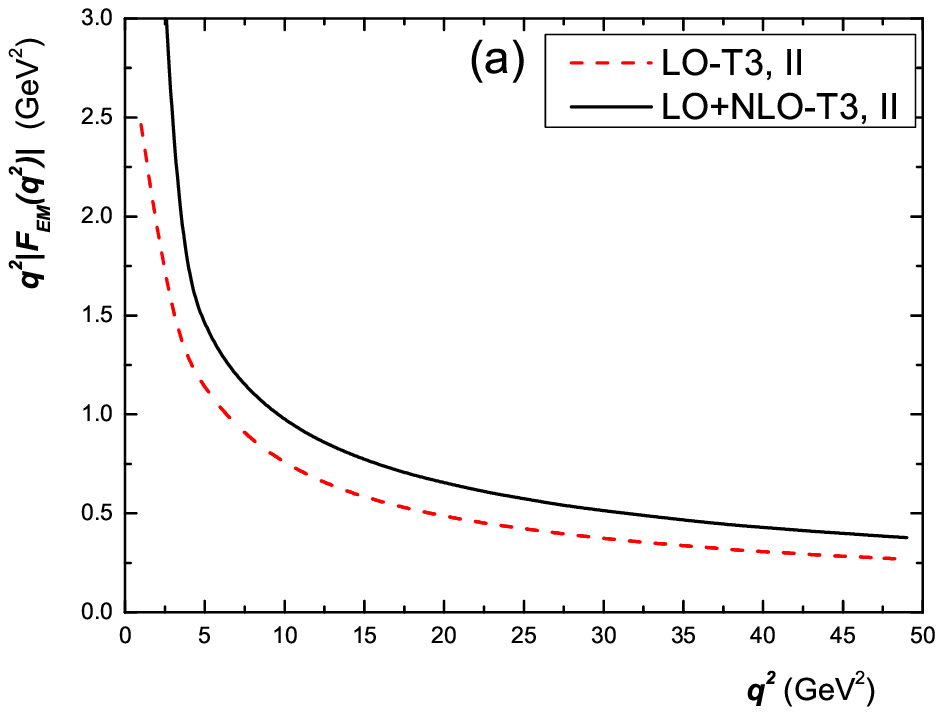}
\includegraphics[width=0.48\textwidth]{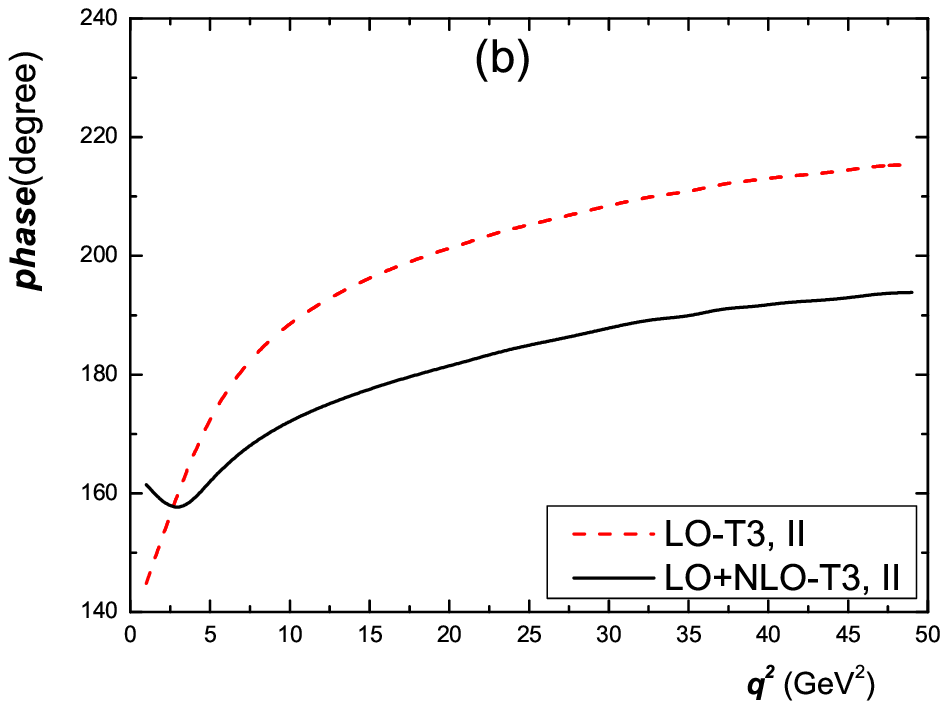}
\vspace{-0.5cm}
\caption{The pQCD predictions for the magnitude and strong phase of time-like pion EM form factors
induced by the twist-3 DAs $\phi^{P,T}$.
The Rome symbol ``II" refers to the form factors calculated in double-b convolution formula
as described in Eq.~(\ref{eq:lotff-double}).}
\label{fig:fig3}
\end{figure*}
%%------------------------------------------------------------------------

%-----------------------------------------------------------------------
\begin{figure*}
\centering
\vspace{-0.5cm}
\includegraphics[width=0.48\textwidth]{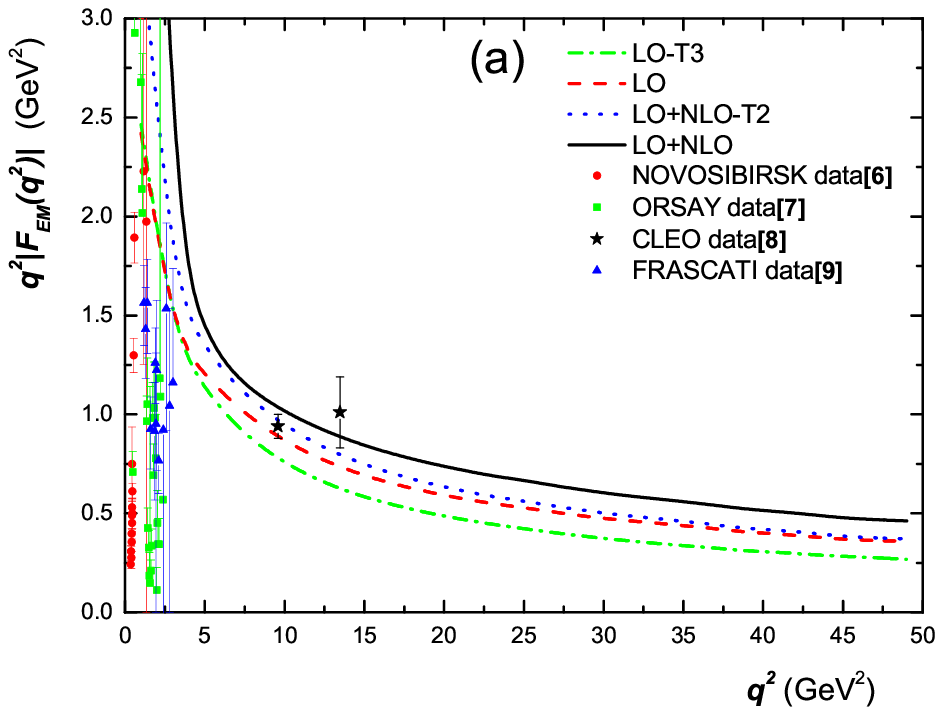}
\includegraphics[width=0.48\textwidth]{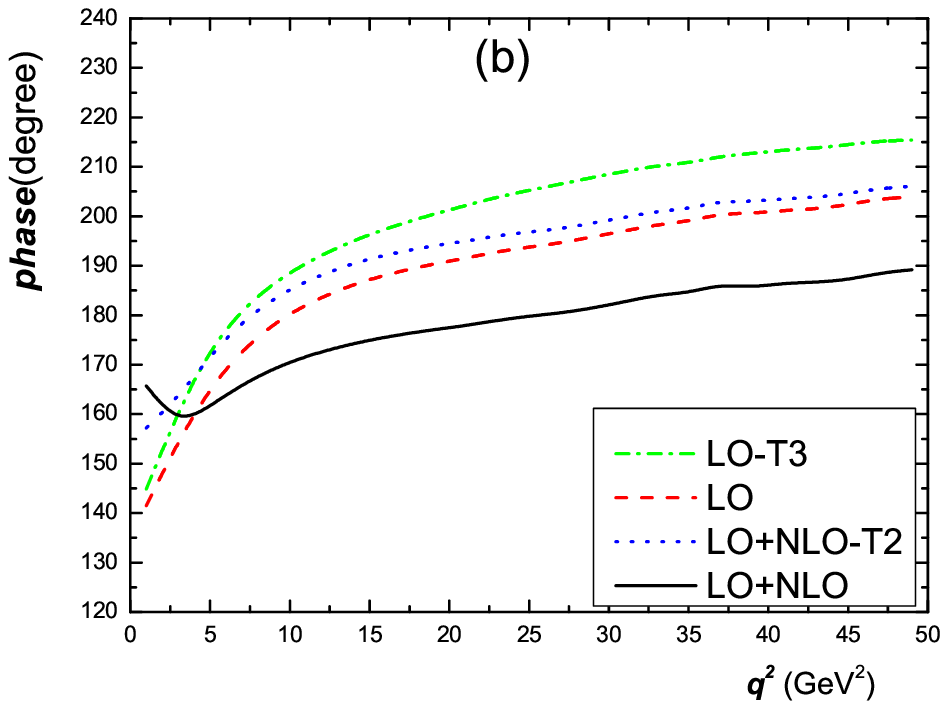}
\vspace{-0.5cm}
\caption{The pQCD predictions for the magnitude and strong phase of time-like pion EM
form factor as escribed in Eqs.~(\ref{eq:lotff-double},\ref{eq:nlotff-double}) at LO and NLO level,
As a comparison, those currently available measured values
\cite{plb073-226,plb220-321,prl95-261803,jpg29-A1} for fixed $q^2$ are also shown in Fig.~4(a).}
\label{fig:fig4}
\end{figure*}
%%------------------------------------------------------------------------

The LO and NLO pQCD predictions for the magnitude and strong phase of time-like pion
EM form factor from the twist-2 and twist-3 DAs are illustrated in Fig.~\ref{fig:fig2}
and Fig.~\ref{fig:fig3} respectively.
By summing up the different twists' contributions, the total pQCD prediction for
these physical quantities are shown in Fig.~\ref{fig:fig4}.
From Fig.~\ref{fig:fig2}, Fig.~\ref{fig:fig3} and Fig.~\ref{fig:fig4}, one can see the
following points:
\begin{enumerate}
\item[(1)]
For the LO form factor induced by the twist-2 DAs, the single-b convolution formula is a good
approximation for the region of $q^2 > 30$ GeV$^2$ because the single-b convolution result
is close to the standard double-b convolution result in this $q^2$ region.
Of course, this approximation can be understood by the fact that the internal gluon propagator
carry almost all the strong phase with no end-point singularity for this twist-2 case.

\item[(2)]
The NLO twist-2 correction to the magnitude (strong phase) of the LO twist-2's contribution
is smaller than $25\%$ ($10^o$) in the region of $q^2 > 30$ GeV$^2$.
The NLO twist-3 correction to the magnitude (strong phase) of the LO twist-3's contribution
is smaller than $35\%$ ($20^o$) in the region of $q^2 > 5$ GeV$^2$.

\item[(3)]
At the LO level, because of the high power singularity, the twist-3 contribution is much
larger than  the twist-2 part in our considered region of $1 < q^2 < 49$ GeV$^2$.
So the obvious NLO twist-3 correction can enhance the LO pQCD prediction
and therefore can improve the agreement between the pQCD prediction and the data,
especially in the region of $q^2 > 5$ GeV$^2$.
The NLO correction with the inclusion of both twist-2 and twist-3 contributions can enhance (reduce)
the magnitude ( strong phase) of the LO one by $20\% - 30\%$ ($< 15^o$) in the region of
$q^2 > 5$ GeV$^2$. The NLO pQCD prediction for time-like form factor
therefore become well consistent with the CLEO data in the of region of $ 5 < q^2 < 15 $ GeV$^2$,
as shown explicitly by the solid curve in Fig.~4(a).

\end{enumerate}

Our numerical result at LO is a little smaller than the one in Ref.~\cite{plb718-1351},
since we here used the different input DAs and the different choice of the QCD scale
$\lambda_{QCD}$.
In Ref.~(\cite{plb718-1351}), $\lambda_{QCD}$ is chosen at the fixed value $0.2$ GeV.
In this paper, however, the QCD scale is varying in the transition process according to the internal hard scale,
and $\lambda_{QCD}$ is around $0.25$ GeV here.

% \section{Conclusion}

In this paper, we firstly gave a brief review for the LO time-like and space-like pion EM form factor
evaluated in the $k_T$ factorization theorem, and then we calculated the NLO twist-3 correction
to the LO time-like pion EM form factor by making the analytic continuation of the NLO twist-3
space-like correction for the corresponding space-like form factor,
and finally we made the numerical calculations for the time-like pion EM form factor
with the inclusion of the NLO twist-2 and twist-3 corrections.

From the numerical results about the LO and NLO pQCD predictions for the
time-like pion EM form factor, we found that:
\begin{enumerate}
\item[(i)]
The LO twist-3 contribution is much larger than the twist-2 one
since the high power end-point singularity;

\item[(ii)]
The NLO twist-2 correction to the LO twist-2 contribution for the magnitude (phase)
is lower than $25\%$ ($10^o$) of the LO form factor in the region of $q^2 > 30$ GeV$^2$.
The NLO twist-3 correction to the LO twist-3 contribution for the magnitude (phase)
of the LO form factor is lower than $35\%$ ($10^o$) in the region of $q^2 > 5$ GeV$^2$.

\item[(iii)]
The total NLO correction with the inclusion of both the twist-2 and twist-3
contributions can enhance (reduce) the magnitude (phase)
of the LO form factor by $20\% - 30\%$ ($< 15^o$) in the region of $q^2 > 5$ GeV$^2$,
and consequently the NLO pQCD prediction for the studied pion EM form factor become
well consistent with the CLEO data.

\end{enumerate}

\section{Acknowledement}

The authors would like to thank Hsiang-nan Li and Cai-Dian L\"u for long term
collaborations and valuable discussions,
and thank Hao-Chung Hu for very useful discussion.
This work is supported by the National Natural Science Foundation of China under
Grant No.10975074 and 11235005.

%%---------------------------------------------------------------------------------------


\begin{thebibliography}{99}


\bibitem{prl43-545}
G.P.~Lepage and S.J.~Brodsky, \prl{\bf 43}, 545(1979);~\prd{\bf 22}, 2157(1980).
%Exclusive processes in quantum chromodynamics: the form factor of baryons at large momentum transfer%
%Exclusice processes in perturbative Quantum chromodynamics%


\bibitem{PRD9-1229}
C.J.~Bebek, et al., (Harvard $\&$ Cornell), \prd{\bf 9}, 1229(1974);~\prd{\bf 13}, 25(1976);~\prd{\bf 17}, 1693(1978).
%Further measurements of forward-charged-pion electroproduction at large $k^2$%
%Deter~fnation of the pion form factor up to $Q^2= 4 Gev^2$ from single-charged-pion electroprofluction%
%Electroproduction of single pions at low $\epsilon$ and a measurement of the pion form factor up to $Q^2 = 10 Gev^2$%

\bibitem{npb137-294}
H.~Ackermann, et al., (DESY Collaboration), \npb{\bf 137}, 294(1978); \non
P.~Brauel, et al., (DESY Collaboration), \zpc{\bf 3}, 101(1979).
%Determination of the longitudinal and the transverse part in \pi^+ electroproduction%
%Electroproduction of $\pi^+n, \pi^-p and K^+\Lambda, K^+\Sigma^0$ Final States Above the Resonance Region%

\bibitem{prl97-192001}
T.~Horn, et al.,~(Jefferson Lab $F_{\pi}$ Collaboration), \prl{\bf 97}, 192001(2006); \non
V.~Tadevosyan, et al.,~(Jefferson Lab $F_{\pi}$ Collaboration), \prc{\bf 75}, 055205(2007).
%Determination of the Pion Charge Form Factor at $Q^2=1:60$ and 2:45 $(GeV/)^2$%
%Determination of the pion charge form factor for $Q^2=0.60¨C1.60 GeV^2$%


\bibitem{prd8-92}
C.N.~Brown, et al., ~(Cyclotron Laboratory at Harvard University), \prd{\bf 8}, 92(1973).
%Coincidence Electroprodnction of Charged Pions and the Pion Form Factor%

\bibitem{plb073-226}
G.K.~Varma and L.~Zamick,~(NOVOSIBIRSK-VEPP-2M-OLYA), \plb{\bf 073}, 226(1978); \non
L.M.~Barkov, et al., ~(NOVOSIBIRSK-VEPP-2M-OLYA), \npb{\bf 256}, 365(1985).
%Charge symmetry considerations in determining neutron radiu in nuclei%
%Electromagnetic pion form factor in the timelike region%

\bibitem{plb220-321}
Ulf-G.~Mei{\ss}ner, ~(ORSAY-DCI-DM2), \plb{\bf 220}, 321(1989).
%Chairity symmetry and medium modifications of nucleon properties%

\bibitem{prl95-261803}
T.K.~Pedlar, et al.,~(CLEO Collaboration), \prl{\bf 95}, 261803(2005).
%Precision Measurements of the Timelike Electromagnetic Form Factors of Pion, Kaon, and Proton%

\bibitem{jpg29-A1}
M.R.~Whalley, \jpg{\bf 29}, A1(2003).
%A compilation of data on hadronic total cross sections in $e^+e^- interactions%


\bibitem{prl65-1717}
C.E.~Carlson and J.~Milana, \prl{\bf 65}, 1717(1990).
%Difficulty in Determining the Pion Form Factor at High $Q^2$%

\bibitem{prd61-073004}
V.M.~Braun, A.~Khodjamirian and M.~Maul, \prd{\bf 61}, 073004(2000).
%Pion form factor in QCD at intermediate momentum transfers%

\bibitem{jhep07-008}
C.~Coriano, H.N.~Li and C.~Savkli, \jhep{\bf 07}, 008(1998).
%Exclusive processes at intermediate energy, quark-hadron duality and the transition to perturbative QCD%

\bibitem{plb094-245}
A.V.~Efrefov and A.V.~Radyushkin, \plb{\bf 094}, 245(1980).
%Factorization and the asymptotic behaviour of pion form factor in QCD%

\bibitem{prd60-074004}
B.~Melic, B.~Nizic and K.~Passek, \prd{\bf 60}, 074004(1999).
%Complete next-to-leading order perturbative QCD prediction for the pion electromagnetic form factor%

\bibitem{prd70-033014}
A.P.~Bakulev, K.~Passek-Kumericki, W.~Schroers and N.G.~Stefanis, \prd{\bf 70}, 033014(2004).
%Pion form factor in QCD: From nonlocal condensates to next-to-leading-order analytic perturbation theory%

\bibitem{prd79-034015}
U.~Raha and A.~Aste, \prd{\bf 79}, 034015(2009).
%Electromagnetic pion and kaon form factors in light-cone resummed perturbative QCD%

\bibitem{plb115-410}
V.A.~Nesterenko and A.V.~Radyushkin, \plb{\bf 115}, 410(1982).
%Sum rules and the pion form factor in QCD%




\bibitem{plb316-546}
P.~Kroll, Th.~Pilsner, M.~Sch\"{u}rmann and W.~Schweiger, \plb{\bf 316}, 546(1993).
%On exclusive raction in the timelike region%

\bibitem{prd47-R3690}
R.~Kahler and J.~Milana, \prd{\bf 47}, R3690(1993); ~ J.~Milana, S.~Nussinov and M. G. Olsson, \prl{\bf 71}, 2533(1993).
%Determining the pion form factor at high $Q^2%
%Does $J\Psi \to \pi^+\pi^-$ Fix the Electromagnetic Form Factor $F_{\pi}(t)$ at $t=M^2_{J\Psi} ?%

\bibitem{prd51-15}
T.~Gousset and B.~Pire, \prd{\bf 51}, 15(1995).
%Timelike form factors at high energy%

\bibitem{prd62-113001}
A.P.~Bakulev, A.V.~Radyushkin and N.G.~Stefanis, \prd{\bf 62}, 113001(2000).
%Form factors and QCD in spacelike and timelike regions%

\bibitem{prd77-113004}
H.M.~Choi and C.R.~Ji, \prd{\bf 77}, 113004(2008).
%Conformal symmetry and pion form factor: Space- and timelike region%

\bibitem{prd82-114012}
U.~Raha and H.~Kohyama, \prd{\bf 82}, 114012(2010).
%Space- and timelike electromagnetic kaon form factors%

\bibitem{plb693-102}
J.W.~Chen, H.~Kohyama, K.~Ohnishi, U.~Raha and Y.L.~Shen, \plb{\bf 93}, 102(2010).
%Space- and time-like electromagnetic pion form factors in light-cone pQCD%




\bibitem{prd47-3875}
T.~Hyer, \prd{\bf 47}, 3875(1993).
%Sudakov effects in p\bar{p} annihilation%

\bibitem{prd66-094010}
H.N.~Li, \prd{ \bf66}, 094010(2002);~\plb {\bf 555}, 197 (2003).
%Threshold resummation for exclusive B meson decays%
%Threshold resummation for nonleptonic B meson decays%

\bibitem{kt theorem}
S.J.~Brodsky and G.R.~Farrar, \prl{\bf 31}, 1153(1973);~
J.~Botts and G.~Sterman, \npb{\bf 325}, 62 (1989);~
S.J.~Brodsky, C.R.~Ji, A.~Pang and D.G.~Robertson, \prd{\bf 57}, 245(1998)~
H.N.~Li and G.~Sterman, \npb {\bf 381}, 129 (1992).
%Scaling laws at largr transverse momentum%
%Hard Elastic Scattering in QCD: Leading Behavior%
%Optimal renormalization scale and scheme for exclusive processes%
%The perturbative pion form factor with Sudakov suppression%
T.~Huang and Q.X. Shen, \zpc {\bf 50}, 139 (1991).
%The applicability of perturbative QCD to the pion form factor and the pion wavefunction%


\bibitem{prd67-034001}
M.~Nagashima and H.N.~Li, \prd{\bf 67}, 034001(2003).
%kT factorization of exclusive processes%

\bibitem{prd52-5358}
F.G.~Cao, T. Huang and C.W.~Luo, \prd {\bf 52}, 5358(1995);~
Z.T.~Wei and M.Z.~Yang, \prd{\bf 67}, 094013(2003).
%Re-examination of the perturbative pion form factor with Sudakov suppression%
%Phenomenological study of Sudakov effects in the pion form factor%

\bibitem{prd84-034018}
Y.C.~Chen and H.N.~Li, \prd{\bf 84}, 034018(2011);~\plb{\bf 712}, 63(2012).
%Three-parton contribution to pion form factor in kT factorization%
%Three-parton contribution to the B¡ú¦Ð form factors in kT factorization%

\bibitem{prd83-054029}
H.N.~Li, Y.L.~Shen, Y.M.~Wang and H.~Zou, \prd{\bf 83}, 054029(2011).
%Next-to-leading-order correction to pion form factor in kT factorization%

\bibitem{plb718-1351}
H.C.~Hu and H.N.~Li, \plb{\bf 718}, 1351(2013).
%Next-to-leading-order time-like pion form factors in kT factorization%

\bibitem{prd63-074009}
C.D.~L\"{u}, K.~Ukai and M.Z.~Yang, \prd {\bf 63}, 074009 (2001);~
T.~Kurimoto, H.N.~Li, and A.I.~Sanda, \prd {\bf 65}, 014007 (2001);~
T.~Huang and X.G.~Wu, \prd{\bf 70}, 093013(2004).
%Branching ratio and CP violation of B ---> pi pi decays in perturbative QCD approach%
%Leading-power contributions to $B \to \pi \rho$ transition form factors%
%Model for the twist-3 wave function of the pion and its contribution to the pion form factor%

\bibitem{prd89-054015}
S.~Cheng, Y.Y.~Fan and Z.J.~Xiao, \prd{\bf 89}, 054015(2014);~
S.~Cheng, Y.Y.~Fan, X.~Yu, C.D,~L\"{u} and Z.J.~Xiao, \prd {\bf 89}, 094004 (2014).
%NLO twist-3 contribution to the pion electromagnetic form factors in kT factorization%
%Next-to-leading-order correction to $B \to \pi$ form factors at twist-3 in $k_{T}$ factorization%

\bibitem{npb2015}
S.~Cheng, Y.L.~Zhang and Z.J.~Xiao, The NLO contributions to the scalar pion form factors and the
${\cal O}(\alpha_s^2)$ annihilation corrections to the $B \to \pi \pi$ , Nuclear Physics B(2015).
http://dx.doi.org/10.1016/j.nuclphysb.2015.04.021
%The NLO contributions to the scalar pion form factors and the $mathcl(O)(\alphas^2_s)$ annihilation corrections to the $B \to \pi \pi$ decays%


\bibitem{prd49-2525}  %% --------------- 36
D. Muller, \prd{\bf 49}, 2525(1993);~\prd{\bf 51}, 3855(1994).
%Conformal constraints and the evolution of the nonsinglet meson distribution amplitude%
%Evolution of the pion distribution amplitude in next-to-leading order%

\bibitem{jhep01-010}
P.~Ball, \jhep{\bf 01}, 010(1998);
%Theoretical Update of Pseudoscalar Meson Distribution Amplitudes of Higher Twist: The Nonsinglet Case%
P.~Ball, V.M.~Braun, Y.~Koike, and K.~Tanaka, \npb{\bf 529}, 323 (1998).
%Higher twist distribution amplitudes of vector mesons in QCD: Formalism and twist - three distributions%

\bibitem{jhep05-004}
P.~Ball, V.M.~Braun and A.~Lenz, \jhep{\bf 05}, 004(2006).
%Higher-twist distribution amplitudes of the K meson in QCD%

\bibitem{prd71-014015}
P.~Ball and R.~Zwicky, \prd{\bf 71}, 014015(2005).
%New results on B \to \pi; K; \eta decay form factors from light-cone sum rules%


\end{thebibliography}
\end{document}